\documentclass[12pt]{article}
\usepackage{amsmath, amssymb}
\usepackage{amsthm}
\usepackage{graphicx}
\usepackage{enumerate}
\usepackage{url} % not crucial - just used below for the URL 

\usepackage[utf8]{inputenc} % set input encoding (not needed with XeLaTeX)
\usepackage[english]{babel}
\usepackage{csquotes}

\newcommand{\blind}{1}

\usepackage{booktabs} % for much better looking tables
\usepackage{tabularray} % tables
\usepackage{tablefootnote}
\usepackage{float} % positioning of tables
\floatstyle{plaintop}
\restylefloat{table}

\IfFileExists{xurl.sty}{\usepackage{xurl}}{} % add URL line breaks if available

\usepackage{graphicx} % support the \includegraphics command and options
\usepackage[outdir=./]{epstopdf}

\usepackage{algorithm}
\usepackage{algpseudocode}

\usepackage{multirow} % Create tables spanning multiple rows
\usepackage{array} % for better arrays (eg matrices) in maths
\usepackage{etoolbox} % Footnotes
\usepackage{paralist} % very flexible & customisable lists (eg. enumerate/itemize, etc.)
\usepackage{verbatim} % adds environment for commenting out blocks of text & for better verbatim
\usepackage{subfig} % make it possible to include more than one captioned figure/table in a single float
\usepackage[colorlinks=true,citecolor=blue,linkcolor=blue]{hyperref} % To link in website links that are clickable

\usepackage{tikz} % To insert figures
\usepackage{xcolor,color}
\usetikzlibrary{fit}

\renewcommand{\labelenumii}{\textbf{A\arabic{enumii}}}

\theoremstyle{remark}
\newtheorem{remark}{Remark}

\theoremstyle{plain}
\newtheorem{theorem}{Theorem}

\newtheorem{lemma}{Lemma}

\usepackage[
authordate,
giveninits=true,
uniquename=false,
backend=biber,
natbib=true,
doi=only,
isbn=false,
url=false,
maxcitenames=2,
noibid
]{biblatex-chicago}

\NewBibliographyString{available}
\NewBibliographyString{urlwhen}

\DefineBibliographyStrings{english}{%
  available = {Available at},
}

\DefineBibliographyStrings{english}{%
  urlwhen = {accessed},
}

\begin{document}

\def\spacingset#1{\renewcommand{\baselinestretch}%
{#1}\small\normalsize} \spacingset{1}

\if1\blind
{
  \title{\bf Inference for Non-Probability Samples Using Error-Prone Auxiliary Information}
  \author{Lyndon Ang\thanks{
    The views expressed in this paper are those of the authors and do not necessarily represent the views of the Australian Bureau of Statistics. Where quoted or used, they should be attributed clearly to the authors.  The first author was supported by funding from the Sir Roland Wilson Foundation and the Australian Bureau of Statistics}\hspace{.2cm}\\
    College of Business and Economics, Australian National University\\
    Robert Clark\\
    College of Business and Economics, Australian National University\\
    Bronwyn Loong\\
    College of Business and Economics, Australian National University\\
    and \\
    Anders Holmberg\\
    Australian Bureau of Statistics}
  \maketitle
} \fi

\if0\blind
{
  \bigskip
  \bigskip
  \bigskip
  \begin{center}
    {\LARGE\bf Inference for Non-Probability Samples Using Error-Prone Auxiliary Information}
\end{center}
  \medskip
} \fi

\bigskip
\begin{abstract}

    Non-probability samples are a potential source of rich information for making statistical inferences.  However, these datasets often suffer from quality issues, in particular selection bias.  The data from a non-probability sample may be combined with auxiliary information from the population or from a reference probability sample to address this bias.  The auxiliary information present on these other sources is generally considered to be error-free, however this is not always the case.  In this paper, we present a novel semiparametric approach to estimate the participation probabilities of units on a non-probability sample while accounting for error-prone auxiliary information in the reference data source.  The approach combines a logistic regression model for the participation probabilities with a nonparametric kernel regression model for the error-prone auxiliary variables.  Through a simulation study and real-world data analysis we show that our proposed approach effectively mitigates selection bias and accounts for measurement error, while estimates from methods which ignore the measurement error still exhibit some bias.  The consistency of the Inverse Probability Weighted estimator produced using the participation probabilities derived from our approach is demonstrated, and a bootstrap variance approach for the estimator is described.

\end{abstract}

\noindent%
\vfill

%%%%%%%%%%%%%%%%%%%%%%%%%%%% Main body of paper

\newpage

\spacingset{1.8} % DON'T change the spacing!
% Paragraph indent
\setlength{\parindent}{0pt}
\setlength{\parskip}{6pt plus 2pt minus 1pt}

\section{Introduction}

Spurred on by declining response rates and increasing costs of running probability-based surveys, there has in recent years been renewed interest in utilising non-probability data to make statistical inferences (\cite{baker.etal_2013}, \cite{ang.etal_2025}).  Non-probability data may provide a rich source of information, but they can also suffer from quality issues.  Selection bias is a particular concern in these datasets, whereby some subgroups of the population are more or less likely to be included in the data.  The selection mechanism for inclusion in these datasets is generally not known.

Various methods have been developed to address selection bias in non-probability data; see for example \citet{zhang_2019}, \citet{rao_2021} and \citet{wu_2022a}.  These methods generally assume that some auxiliary information is available from another data source.  The other data source may be population-based, for example from a census, or may be from a probability survey, also known as a reference sample.  Methods also vary depending on the data items that are present on the reference sample versus the non-probability sample.  We note that the situation we are describing differs somewhat from the classical survey definition of auxiliary variables (see for example \citealt{cochran_1977} and \citealt{sarndal.etal_1992}) which assumes that auxiliary information is available for the full population.  Additionally, this paper describes a situation where the non-probability data source is without error while the reference sample data has some measurement error, which is also a departure from classical survey theory.  These two points reflect the evolving role of non-survey data to produce statistics.

In this paper, we focus on the so-called quasi-randomisation approach for addressing selection bias, in the case where a data item of interest is available on the non-probability sample while auxiliary data items are available on both the non-probability sample and the reference sample.  In this approach, the missingness in the non-probability sample is assumed to be governed by an unknown selection mechanism that can be modelled using the available auxiliary data.  The model can be applied to units in the non-probability sample to estimate their propensities of inclusion in the non-probability sample, also known as participation probabilities.  The inverse of these participation probabilities are then used as weights to form an Inverse Probability Weighted (IPW) estimator for the non-probability sample to estimate finite population quantities.

Methods for addressing selection bias generally make the simplifying assumption that the data contained in the reference and non-probability samples do not have measurement error.  Measurement error is defined broadly in this paper as a misalignment between the true value of a variable and the value actually captured by the sample.  Measurement error may arise, for example, if a reference sample is from one or two years ago relative to the non-probability sample, and so some of the variables on that dataset are out of date.  Other causes of measurement error may include, for example, the use of concept definitions which differ from what is intended, input error, or rounding error.

We are particularly interested in the specific scenario where measurement error exists in some of the auxiliary variables present in the reference sample, while the data items on the non-probability sample are error-free.  This situation is likely to become more and more common as the frequency of integration of probability samples with non-probability data increases.  Two examples where this scenario applies are:
\begin{itemize}
    \item Out-of-date frame variables - In the business survey context, administrative business taxation data is often brought into the National Statistical Office (NSO) as auxiliary information on the population frame for business surveys.  This data may only be updated periodically and hence be out-of-date when used as auxiliary information.  This example will be the focus of the real-world data simulation in Section \ref{sec5:real.data}.
    \item Different concept definitions - Definitions of certain concepts may differ between the non-probability sample and the reference sample it is being integrated with.  For example, an administrative dataset may define gross income in a certain way, but the gross income data on a reference sample has a slightly different set of items that it includes in its definition of gross income.
\end{itemize}

In this paper, we will sometimes refer to the error-prone auxiliary variables as covariate measurement error given that the auxiliary variables are used as covariates in a model to predict participation probabilities.  Measurement error in the auxiliary variables will potentially introduce attenuation bias when estimating model parameters (\cite{carroll.etal_2006}), and this so-called attenuation bias can lead to inaccurately estimated participation probabilities for the non-probability sample units, and thus ineffective correction of selection bias.  Note that our setup is somewhat complicated by the fact that in our case the dependent variable that we are modelling is a probability of being selected in the non-probability dataset, and the auxiliary variables are available without error on the non-probability dataset but with error on the reference sample units not in the non-probability dataset.  As such, the effect of measurement error may not be a simple attenuation.

There has been a large body of work undertaken on the general topic of measurement error.  \citet{fuller_1987} and \citet{carroll.etal_2006} produced monographs examining the effect of measurement error on linear and non-linear models, while the monograph published by \citet{yi_2017} looked at analysing measurement error and misclassification problems arising in the medical research and epidemiological fields.  \citet{stefanski_2000} provides a brief introduction to measurement error problems and reviews some key contributions to the topic.

The literature on addressing measurement error issues specifically for non-probability samples is still relatively sparse.  \citet{valliant.dever_2011} pointed out that covariate measurement error may lead to biased estimation of participation probabilities.  \citet{schenkel.zhang_2022} developed a modelling approach to adjust for misclassification in a target variable on an administrative dataset, using a second classifier from another source that is also subject to misclassification.  \citet{dharma.etal_2025} examined the case when a data item collected in the full reference sample is used to identify a subgroup of interest corresponding to the target population of a non-probability sample, and that data item could be misclassified - specifically, that there are some units that should be included in the subgroup but are not.  The authors proposed to estimate the probability of misclassification and then generate a predicted classification to be used to define the subgroup.

The calibration approach proposed in \citet{kim.tam_2021} for integrating probability and non-probability samples allows for measurement error in the target variable on either of the two samples.  Given that their approach applies to the case when both target and auxiliary variables are available on both samples, it is possible to fit a measurement error model using the overlapping units in the reference and non-probability samples.  The fitted model can then be used to adjust the mis-measured data on the reference sample.  In \citet{ang.etal_2025}, an empirical study was conducted to examine the impact of measurement error in the target variable on estimates of totals, and the effectiveness of using a measurement error model to correct for those errors.

This paper makes a major contribution to existing literature in several ways.  To our knowledge, we are the first to explore the effects of error-prone auxiliary variables on the reference sample which is to be combined with a non-probability sample to model participation probabilities.  We present a semiparametric approach to simultaneously address this covariate measurement error in the reference sample and selection bias in the non-probability sample, when it is possible to integrate the reference sample and the non-probability sample.  In this scenario, information about the true values of the error-prone auxiliary variables is available for the units that are in both the non-probability and reference samples, making it feasible to model the measurement error.  It is assumed that the mis-measured auxiliary variables do not influence participation probabilities over and above the correctly measured versions of those variables. Under this assumption, a parametric model for the participation probabilities is combined with a nonparametric kernel regression model for the mis-measured auxiliary variables to obtain a semiparametric maximum likelihood estimator for the parameters of the participation probability model.  Of note, we show that the proposed framework can also be used to correct selection bias in a non-probability sample that has nonignorable missingness issues.  Through two simulation studies using a synthetic population and real business data, we demonstrate that ignoring covariate measurement error will lead to biased results, while our proposed approach effectively mitigates the bias.

The rest of the paper is structured as follows.  Section \ref{setup} outlines the basic setup.  Section \ref{sec5:method} introduces the proposed semiparametric framework for correcting measurement error and selection bias.  In Section \ref{sec5:consistency.variance} it is shown that the proposed semi-parametric estimating function is consistent, and a bootstrap approach for estimating variance is outlined.  In Section \ref{sec5:simstudy} the performance of the proposed approach is examined through a simulation study.  The approach is examined using real-world data in Section \ref{sec5:real.data}.  Section \ref{sec5:cov.error.B} briefly discusses how a similar approach may be employed to address measurement error in the auxiliary variables in the non-probability sample.  Finally, Section \ref{sec5:conclusion} provides some concluding remarks.

\section{Basic Setup} \label{setup}

%%% For assumptions
\renewcommand{\labelenumii}{\textbf{A\arabic{enumii}}}

Let $\mathcal{U}$ be a finite population of size $N$.  For each unit $i$ in the population, define values $(y_i, \boldsymbol{x}_i, \boldsymbol{z}_i)^T$ for a data item of interest $y$ and auxiliary data items $(\boldsymbol{x}, \boldsymbol{z})$.  $\boldsymbol{x}$ may suffer from measurement error, while $\boldsymbol{z}$ is always free from measurement error.  Denote the mis-measured version of $\boldsymbol{x}$ using $\boldsymbol{x}^*$.  This paper focuses on estimating the population mean $\mu_y = N^{-1}\sum_{i=1}^{N}y_i$, but the approach also applies to population totals.

A reference probability sample $A$ of size $n_A$ is drawn from the population.  $A$ contains information for $\boldsymbol{z}$ and $\boldsymbol{x}^*$, and does not contain information for $y$.  Define $\pi^A_i=P(i \in A|\mathcal{U})$ as the inclusion probability for unit $i$ being in the probability sample, and $d^A_i=1/\pi^A_i$ as the survey design weight for $i \in A$.  The $\pi^A_i$ and $d^A_i$ values are known from the sample design.

A non-probability sample $B$ is also available from the population.  $B$ contains information for $\boldsymbol{z}$ and $\boldsymbol{x}$, as well as the variable of interest $y$.  Let $\delta_i=I(i \in B)$ be an indicator variable for unit $i$ being included in the sample $B$.  The non-probability sample size is $N_B=\sum_{i=1}^{N}{\delta_i}$.  In contrast to $\pi^A_i$ in the reference probability sample, $\pi^B_i=P(\delta_i=1|\mathcal{U})$ is unknown and needs to be estimated.  Define $C=\mathcal{U} \setminus B$ as the units in the population not included in $B$.  Let $n_{AB}$ be the size of the overlapping sample $A \cap B$.

A number of assumptions are made regarding the selection mechanism for $B$.  These assumptions are:
\begin{enumerate}
    %\addtolength\itemsep{-4mm}
	\item[]
    \begin{enumerate}
        %\addtolength\itemsep{-4mm}
        \item \textit{Ignorability}: Conditional on the set of covariates $\boldsymbol{x}_i$, $\delta_i$ and $y_i$ are independent.
        \item \textit{Positivity}:  Conditional on $(\boldsymbol{x}_i,\boldsymbol{z}_i)$, $P(\delta_i=1|\boldsymbol{x}_i, \boldsymbol{z}_i) > 0, \forall i \in \mathcal{U}$
        \item \textit{Independence}: Conditional on $(\boldsymbol{x}_i,\boldsymbol{z}_i)$ and $(\boldsymbol{x}_j,\boldsymbol{z}_j)$, $\delta_i \perp \delta_j, i \neq j, \forall i,j \in \mathcal{U}$
    \end{enumerate}
\end{enumerate}

Ignorability implies that $P(\delta_i=1|\boldsymbol{x},y)=P(\delta_i=1|\boldsymbol{x})$.  In other words, selection into $B$ is ignorable conditional on the covariates $\boldsymbol{x}$.  This assumption is similar to the Missing-At-Random (MAR) scenario of \citet{rubin_1976}.  \citet{andridge.etal_2019} and \citet{boonstra.etal_2021} refer to this type of selection process as Selection At Random (SAR).  When selection into $B$ is influenced by $y$, then we instead have Selection Not At Random (SNAR), akin to the Missing-Not-At-Random (MNAR)/Not-Missing-At-Random (NMAR) scenario.  Later in this paper, it will be shown that the proposed measurement error framework can also be applied to resolve SNAR issues in the non-probability sample.

The positivity assumption \textbf{A2} states that conditional on $(\boldsymbol{x}, \boldsymbol{z})$ every unit in the population has a non-zero chance of inclusion in $B$.  This may not always hold - for example, in a big dataset generated by a social media platform, only those persons who are members of the platform have a chance of inclusion in the big dataset.  The independence assumption \textbf{A3} facilitates forming a likelihood function (see for example \citealt{wu_2023}).

The following assumption will also be made in relation to the measurement error and selection mechanisms:
\begin{enumerate}
    %\addtolength\itemsep{-4mm}
	\item[]
    \begin{enumerate}
        %\addtolength\itemsep{-4mm}
        \setcounter{enumii}{3}
        \item \textit{Non-differential measurement error}: Conditional on the set of covariates $(\boldsymbol{x}_i, \boldsymbol{z}_i)$, $\delta_i$ and $\boldsymbol{x}^*_i$ are independent
    \end{enumerate}
\end{enumerate}

Assumption \textbf{A4} says that the mis-measured $\boldsymbol{x}^*$ does not give any further information for inference about the participation probabilities given the true covariates.  In other words, $P(\delta|\boldsymbol{x}^*, \boldsymbol{x}, \boldsymbol{z}) = P(\delta|\boldsymbol{x}, \boldsymbol{z})$.

Finally, it is assumed that there is some overlap between $A$ and $B$, and it is possible to accurately link the units in $A$ and $B$.  The overlap assumption is reasonable if $A$ and $B$ are not too small, or if the sample design of $A$ facilitates some overlap with units in $B$.  The linkage assumption may be satisfied if there exists a common unit identifier available on $B$ and the reference sample $A$, or if it is possible to accurately link records in $A$ and $B$ using a common set of linking variables.

\section{Proposed Method} \label{sec5:method}

Recall that the end goal in this paper is to produce an estimate of the population mean using $B$.  This can be achieved by producing $\hat{\pi}^B_i$ and feeding it into an Inverse Probability Weighted (IPW) estimator such as
\begin{equation} \label{eqn5:IPW}
    \hat{\mu}_{IPW} = \frac{1}{\hat{N}_B}\sum_{i \in B}\frac{y_i}{\hat{\pi}^B_i}
\end{equation}
where $\hat{N}_B = \sum_{i \in B}(\hat{\pi}^B_i)^{-1}$.  The problem addressed in this paper is how best to produce $\hat{\pi}^B_i$.

It is assumed that the $\pi^B_i$ are governed by a parametric model, so that $\pi^B = P(\delta = 1| \boldsymbol{x}, \boldsymbol{z}; \boldsymbol{\theta}_0) =  \pi(\boldsymbol{x}, \boldsymbol{z}; \boldsymbol{\theta}_0)$, where $\boldsymbol{\theta}_0$ are the true but unknown model parameters for the selection mechanism.  We first consider the case when $A$ corresponds to a population-level dataset, for instance when it is the population frame or data from a population census.

If $\boldsymbol{x}$ is measured accurately in $A$ and $B$, then the likelihood function for $\{\delta_i, \boldsymbol{x}_i, \boldsymbol{z}_i; i \in B\}$ and $\{\boldsymbol{x}_i, \boldsymbol{z}_i; i \in \mathcal{U} \}$ would be given by
\begin{equation} 
    L_p(\boldsymbol{\theta}) = \prod_{i \in \mathcal{U}} \pi(\boldsymbol{x}_i, \boldsymbol{z}_i; \boldsymbol{\theta})^{\delta_i}(1-\pi(\boldsymbol{x}_i, \boldsymbol{z}_i; \boldsymbol{\theta}))^{(1-\delta_i)} \nonumber
\end{equation}

and the corresponding log-likelihood function is
\begin{equation} \label{eqn:loglikelihood_full}   
    l_p(\boldsymbol{\theta}) = \sum_{i \in \mathcal{U}}\delta_i \text{log}\pi(\boldsymbol{x}_i, \boldsymbol{z}_i; \boldsymbol{\theta}) + \sum_{i \in \mathcal{U}}(1 - \delta_i)\text{log}(1 - \pi(\boldsymbol{x}_i, \boldsymbol{z}_i; \boldsymbol{\theta}))
\end{equation}

\begin{remark} \label{rem:oe.likelihood}
    When $\boldsymbol{x}$ is not available for the full population, the second term in (\ref{eqn:loglikelihood_full}) is not observed.  Assuming that it is possible to accurately identify which units in $A$ are also in $B$, this term can be estimated using $A \setminus B$.  The resulting pseudo-likelihood was termed the Overlap-Excluded (OE) likelihood in \citet{ang.etal_2024a}, and may be maximised to obtain $\hat{\boldsymbol{\theta}}$ for the participation probability model; see \citet{ang.etal_2024a} for further details.  The pseudo-likelihood is given as
    \begin{equation} \label{eqn:oe.loglikelihood}   
        l_{OE}(\boldsymbol{\theta}) = \sum_{i \in B}\text{log}\pi^B_i + \sum_{i \in A \setminus B}d^A_i \text{log}(1 - \pi^B_i)
    \end{equation}
    
\end{remark}

When $\boldsymbol{x}$ is instead measured by its error-prone version $\boldsymbol{x}^*$ in $A$, the likelihood function for $\{\delta_i, \boldsymbol{x}_i, \boldsymbol{z}_i; i \in B\}$ and $\{\boldsymbol{x}^*_i, \boldsymbol{z}_i; i \in \mathcal{U} \}$ becomes
\begin{equation} \label{eqn5:obs.likelihood}  
  L(\boldsymbol{\theta}) = \prod_{i \in \mathcal{U}}  \pi(\boldsymbol{x}_i,\boldsymbol{z}_i;\boldsymbol{\theta})^{\delta_i} \Bigl\{ \int (1-\pi(\boldsymbol{x}_i,\boldsymbol{z}_i;\boldsymbol{\theta}))f(\boldsymbol{x}|\boldsymbol{x}^*_i, \boldsymbol{z}_i)\text{d}\boldsymbol{x} \Bigr \} ^{1-\delta_i}.
\end{equation}

The corresponding log-likelihood function is
\begin{equation} \label{eqn5:obs.loglikelihood}   
  l(\boldsymbol{\theta}) = \sum_{i \in \mathcal{U}} \delta_i \text{log}\pi(\boldsymbol{x}_i, \boldsymbol{z}_i; \boldsymbol{\theta}) + \sum_{i \in \mathcal{U}}(1 - \delta_i)\text{log} \Bigl\{ \int (1 - \pi(\boldsymbol{x}_i, \boldsymbol{z}_i; \boldsymbol{\theta}))f(\boldsymbol{x}|\boldsymbol{x}^*_i, \boldsymbol{z}_i)\text{d}\boldsymbol{x} \Bigr\}
\end{equation}
and, assuming it is possible to interchange differentiation and integration, the score function is given by
\begin{align} 
  s(\boldsymbol{\theta}) &= \frac{\partial l} {\partial \boldsymbol{\theta}} \nonumber \\ 
  &= \sum_{i \in \mathcal{U}} \delta_i \frac{\pi'(\boldsymbol{x}_i, \boldsymbol{z}_i; \boldsymbol{\theta})}{\pi(\boldsymbol{x}_i, \boldsymbol{z}_i; \boldsymbol{\theta})} - \sum_{i \in \mathcal{U}}(1 - \delta_i)\frac{\int \pi'(\boldsymbol{x}_i, \boldsymbol{z}_i; \boldsymbol{\theta})f(\boldsymbol{x} | \boldsymbol{x}^*_i, \boldsymbol{z}_i)\text{d}\boldsymbol{x}}{\int (1 - \pi(\boldsymbol{x}_i, \boldsymbol{z}_i; \boldsymbol{\theta}))f(\boldsymbol{x} | \boldsymbol{x}^*_i, \boldsymbol{z}_i)\text{d}\boldsymbol{x}} \label{eqn5:score}
\end{align}
where $\pi'(\boldsymbol{x}, \boldsymbol{z}; \boldsymbol{\theta}) = \partial\pi(\boldsymbol{x}, \boldsymbol{z}; \boldsymbol{\theta})/\partial\theta$ is the derivative of $\pi(\boldsymbol{x}, \boldsymbol{z}; \boldsymbol{\theta})$ with respect to $\boldsymbol{\theta}$.  Solving the score function requires knowledge of $f(\boldsymbol{x} | \boldsymbol{x}^*_i, \boldsymbol{z}_i)$.  It may be possible to obtain this information through, for example, taking a validation subsample from the original sample $A$.  However, this is not always possible to do, for example due to budget constraints.  However, note that data is available on $(\boldsymbol{x}_i, \boldsymbol{z}_i)$ from $B$, and $(\boldsymbol{x}^*_i, \boldsymbol{z}_i)$ from $A$.  If it is possible to link the data from $A$ and $B$, then one can obtain $(\boldsymbol{x}_i, \boldsymbol{x}^*_i, \boldsymbol{z}_i)$ for records in $A \cap B$.  Thus, one can instead estimate $f(\boldsymbol{x} | \boldsymbol{x}^*_i, \boldsymbol{z}_i, \delta_i = 1) = f_1(\boldsymbol{x} | \boldsymbol{x}^*_i, \boldsymbol{z}_i)$ using $A \cap B$, where $f_1(\cdot) = f(\cdot|\delta_i=1)$.

Using Bayes' Theorem, we have
\begin{align}
    f_1(\boldsymbol{x} | \boldsymbol{x}^*, \boldsymbol{z}) &= f(\boldsymbol{x} | \boldsymbol{x}^*, \boldsymbol{z}, \delta = 1)
    = \frac{P(\delta=1 | \boldsymbol{x}, \boldsymbol{x}^*, \boldsymbol{z})f(\boldsymbol{x}|\boldsymbol{x}^*, \boldsymbol{z})}{P(\delta=1|\boldsymbol{x}^*, \boldsymbol{z})} \nonumber
\end{align}
and assuming that the non-differential error mechanism holds, we have $P(\delta=1|\boldsymbol{x}, \boldsymbol{x}^*, \boldsymbol{z}) = P(\delta=1|\boldsymbol{x}, \boldsymbol{z}) = \pi^B$.  So
\begin{align}
    f(\boldsymbol{x} | \boldsymbol{x}^*, \boldsymbol{z}) = f_1(\boldsymbol{x} | \boldsymbol{x}^*, \boldsymbol{z})\frac{P(\delta=1|\boldsymbol{x}^*, \boldsymbol{z})}{\pi^B}. \label{eqn5:distreln}
\end{align}

Substituting (\ref{eqn5:distreln}) into the score function (\ref{eqn5:score}) gives
\begin{align}
    s(\boldsymbol{\theta}) = &\sum_{i \in \mathcal{U}} \delta_i \frac{\pi'(\boldsymbol{x}_i, \boldsymbol{z}_i; \boldsymbol{\theta})}{\pi(\boldsymbol{x}_i, \boldsymbol{z}_i; \boldsymbol{\theta})} - \nonumber \\
    &\sum_{i \in \mathcal{U}}(1 - \delta_i)\frac{\int \pi'(\boldsymbol{x}, \boldsymbol{z}_i; \boldsymbol{\theta}) f_1(\boldsymbol{x} | \boldsymbol{x}^*_i, \boldsymbol{z}_i){P(\delta_i=1|\boldsymbol{x}^*_i, \boldsymbol{z}_i)}/\pi(\boldsymbol{x}, \boldsymbol{z}_i;\boldsymbol{\theta}) \text{d}\boldsymbol{x}}{\int (1 - \pi(\boldsymbol{x}, \boldsymbol{z}_i; \boldsymbol{\theta}))f_1(\boldsymbol{x} | \boldsymbol{x}^*_i, \boldsymbol{z}_i){P(\delta_i=1|\boldsymbol{x}^*_i, \boldsymbol{z}_i)}/\pi(\boldsymbol{x}, \boldsymbol{z}_i;\boldsymbol{\theta})\text{d}\boldsymbol{x}} \nonumber \\
    =& \sum_{i \in \mathcal{U}} \delta_i \frac{\pi'(\boldsymbol{x}_i, \boldsymbol{z}_i; \boldsymbol{\theta})}{\pi(\boldsymbol{x}_i, \boldsymbol{z}_i; \boldsymbol{\theta})} - \nonumber \\
    &\sum_{i \in \mathcal{U}}(1 - \delta_i)\frac{\int \pi'(\boldsymbol{x}, \boldsymbol{z}_i; \boldsymbol{\theta}) (1 - \pi(\boldsymbol{x}, \boldsymbol{z}_i; \boldsymbol{\theta}_0)))^{-1}O(\boldsymbol{x},\boldsymbol{z}_i)f_1(\boldsymbol{x} | \boldsymbol{x}^*_i, \boldsymbol{z}_i) \text{d}\boldsymbol{x}}{\int O(\boldsymbol{x}, \boldsymbol{z}_i)f_1(\boldsymbol{x} | \boldsymbol{x}^*_i, \boldsymbol{z}_i)\text{d}\boldsymbol{x}} \label{eqn5:score_subf1}
\end{align}
where $O(\boldsymbol{x},\boldsymbol{z}_i) = (1 - \pi(\boldsymbol{x}, \boldsymbol{z}_i))/\pi(\boldsymbol{x}, \boldsymbol{z}_i)$ and the $P(\delta_i = 1|\boldsymbol{x}^*_i, \boldsymbol{z}_i)$ terms can be taken out of the integrals and cancel out.

The integrals may be evaluated by assuming a parametric form for $f_1(\boldsymbol{x}|\boldsymbol{x}^*, \boldsymbol{z})$ (see for example \citealt{riddles.etal_2016}).  However, this approach carries a risk of model misspecification.  Instead, we propose to adopt a nonparametric regression approach similar to that considered in \citet{morikawa.etal_2017}.  Specifically, we take advantage of the fact that the integral terms in (\ref{eqn5:score_subf1}) are conditional expectations.  In particular, they are expectations that are conditional on membership in $B$.  That is,
\begin{align}
    \int(\cdot)f_1(\boldsymbol{x}|\boldsymbol{x}^*, \boldsymbol{z}) = E_1[(\cdot) | \boldsymbol{x}^*, \boldsymbol{z}]. \nonumber
\end{align}

If $A$ comprises the full population, then the Nadaraya-Watson (\citealt{nadaraya_1964}, \citealt{watson_1964}) kernel regression estimator of $E_1[(\cdot) | \boldsymbol{x}^*, \boldsymbol{z}]$ is given by
\begin{align}
    E_1[(\cdot) | \boldsymbol{x}^*, \boldsymbol{z}] = \frac{\sum_{j \in B}K_h(\tilde{\boldsymbol{x}}_i, \tilde{\boldsymbol{x}}_j)(\cdot)}{\sum_{j \in B}K_h(\tilde{\boldsymbol{x}}_i, \tilde{\boldsymbol{x}}_j)} \label{eqn5:kernel.exp}
\end{align}
where $\tilde{\boldsymbol{x}}=(\boldsymbol{x}^*,\boldsymbol{z})$ is a combined vector of covariates, and $K_h$ relates to a given kernel function with bandwidth $h$.  The score function (\ref{eqn5:score_subf1}) is thus estimated by
\begin{align}
    \hat{s}(\boldsymbol{\theta}) &= \sum_{i \in \mathcal{U}} \delta_i \frac{\pi'(\boldsymbol{x}_i, \boldsymbol{z}_i; \boldsymbol{\theta})}{\pi(\boldsymbol{x}_i, \boldsymbol{z}_i; \boldsymbol{\theta})} - \nonumber \\
    &\sum_{i \in \mathcal{U}}(1 - \delta_i)\frac{\sum_{j \in B} \pi'(\boldsymbol{x}_j, \boldsymbol{z}_i; \boldsymbol{\theta}) (1 - \pi(\boldsymbol{x}_j, \boldsymbol{z}_i;\boldsymbol{\theta}))^{-1}O(\boldsymbol{x}_j,\boldsymbol{z}_i)K_h(\tilde{\boldsymbol{x}}_j, \tilde{\boldsymbol{x}}_i)}{\sum_{k \in B} O(\boldsymbol{x}_k, \boldsymbol{z}_i)K_h(\tilde{\boldsymbol{x}}_k, \tilde{\boldsymbol{x}}_i)} \nonumber \\
    &= \sum_{i \in \mathcal{U}} \delta_i \frac{\pi'(\boldsymbol{x}_i, \boldsymbol{z}_i; \boldsymbol{\theta})}{\pi(\boldsymbol{x}_i, \boldsymbol{z}_i; \boldsymbol{\theta})} - \sum_{i \in \mathcal{U}}(1 - \delta_i)\sum_{j \in B} \pi'(\boldsymbol{x}_j, \boldsymbol{z}_i; \boldsymbol{\theta}) (1 - \pi(\boldsymbol{x}_j, \boldsymbol{z}_i;\boldsymbol{\theta}))^{-1}w_{ij} \label{eqn5:score_kr} \\
    &\text{ where } w_{ij}=\frac{O(\boldsymbol{x}_j, \boldsymbol{z}_i)K_h(\tilde{\boldsymbol{x}}_j, \tilde{\boldsymbol{x}}_i)}{\sum_{k \in B} O(\boldsymbol{x}_k, \boldsymbol{z}_i)K_h(\tilde{\boldsymbol{x}}_k, \tilde{\boldsymbol{x}}_i)}. \nonumber
\end{align}

A solution to the score equation $\hat{s}(\boldsymbol{\theta}) = 0$ may be obtained using the EM algorithm, where updated values of $w_{ij}$ and $\hat{\boldsymbol{\theta}}$ are produced in an iterative fashion.  The EM procedure begins with a starting set of values $\boldsymbol{\theta}^{(0)}$, which are fed into $O(\boldsymbol{x}_j, \boldsymbol{z}_i)$ to produce a new set of $w_{ij}$ which are then used to solve the score equation $\hat{s}(\boldsymbol{\theta}) = 0$, for example using the Newton-Raphson algorithm, to obtain the updated set of parameter values $\boldsymbol{\theta}^{(t+1)}$.  The process iterates until convergence.

$\hat{\pi}^B_i$ can be obtained using the values $\hat{\boldsymbol{\theta}}$, and these can then be applied in (\ref{eqn5:IPW}) to produce an estimate of the population mean.

If the reference sample $A$ is a probability sample taken from $\mathcal{U}$, then the score function (\ref{eqn5:score_kr}) cannot be used directly.  Instead, one can replace the second sum over $\mathcal{U}$ by an inverse selection probability estimator over $A$, and also estimate its interior sum over $A \cap B$ as follows:
\begin{align}
    \hat{s}_p(\boldsymbol{\theta}) &= \sum_{i \in \mathcal{U}} \delta_i \frac{\pi'(\boldsymbol{x}_i, \boldsymbol{z}_i; \boldsymbol{\theta})}{\pi(\boldsymbol{x}_i, \boldsymbol{z}_i; \boldsymbol{\theta})} - \sum_{i \in A}d^A_i(1 - \delta_i)\sum_{j \in A \cap B} \pi'(\boldsymbol{x}_j, \boldsymbol{z}_i; \boldsymbol{\theta}) (1 - \pi(\boldsymbol{x}_j, \boldsymbol{z}_i;\boldsymbol{\theta}))^{-1}w_{ij} \label{eqn5:score_kr_sample} \\
    &\text{ where } w_{ij}=\frac{d^A_j O(\boldsymbol{x}_j, \boldsymbol{z}_i)K_h(\tilde{\boldsymbol{x}}_j, \tilde{\boldsymbol{x}}_i)}{\sum_{k \in A \cap B} d^A_k O(\boldsymbol{x}_k, \boldsymbol{z}_i)K_h(\tilde{\boldsymbol{x}}_k, \tilde{\boldsymbol{x}}_i)}. \nonumber
\end{align}

\begin{remark} \label{rem5:snar.framework}
    The framework that has been proposed for addressing covariate measurement error in $A$ is also applicable for the case when the selection mechanism in $B$ is SNAR and $y$ is only collected in $B$.  Using our proposed framework, it is possible to treat the missing $y$ variable in $A$ as an extreme case of measurement error where the measurement of $y$ is missing.  In other words, using the notation in Section \ref{setup}, $\boldsymbol{x}$ now includes the $y$ variable and $\boldsymbol{x}^*$ omits it.  To avoid potential model identifiability problems in the SNAR scenario (see for example \citealt{shao.wang_2016} and \citealt{wang.etal_2014} in the MNAR literature), it is assumed that the set of auxiliary variables $\boldsymbol{z}$ can be separated into two groups  $(\boldsymbol{z}_1, \boldsymbol{z}_2)$, and the participation probability model only depends on $y$ and the variables in $\boldsymbol{z}_1$, that is, $\pi^B = P(\delta=1|y, \boldsymbol{z}_1)$.  The variables $\boldsymbol{z}_2$ and $\delta$ are conditionally independent given $\boldsymbol{z}_1$ and $y$, though $\boldsymbol{z}_2$ may still be related to $y$.  The kernel score function for the SNAR case is then modified slightly to give
    \begin{align}
        \hat{s}_n(\boldsymbol{\theta})&= \sum_{i \in \mathcal{U}} \delta_i \frac{\pi'(\boldsymbol{z}_{1i}, y_i; \boldsymbol{\theta})}{\pi(\boldsymbol{z}_{1i}, y_i; \boldsymbol{\theta})} - \sum_{i \in A} d^A_i (1 - \delta_i)\sum_{j \in A \cap B} \pi'(\boldsymbol{z}_{1i}, y_j; \boldsymbol{\theta}) (1 - \pi(\boldsymbol{z}_{1i}, y_j;\boldsymbol{\theta}))^{-1}w_{ij} \label{eqn5:score_kr_snar} \\
    &\text{ where } w_{ij}=\frac{d^A_jO(\boldsymbol{z}_{1i}, y_j)K_h(\boldsymbol{z}_j, \boldsymbol{z}_i)}{\sum_{k \in A \cap B} d^A_k O(\boldsymbol{z}_{1i}, y_k)K_h(\boldsymbol{z}_k, \boldsymbol{z}_i)} \nonumber
    \end{align}
    where $d^A_i=1$ and $A \cap B = B$ if $A$ comprises the full population.
\end{remark}

The Supplementary Materials provides a detailed derivation of $\hat{s}_n$ from the nonignorable missingness perspective.

\begin{remark}
    Using a similar reasoning to Remark \ref{rem5:snar.framework}, it is possible to apply the proposed framework to the case when some covariates that should be included in the participation probability model are missing from the reference sample $A$.  In this case, suppose that the set of auxiliary variables in the participation probability model should include $(\boldsymbol{x},\boldsymbol{z})$.  The auxiliary variables missing from $A$ but present on $B$ are included in $\boldsymbol{x}$, while $\tilde{\boldsymbol{x}}$ now includes only $\boldsymbol{z}$.  Model identifiability issues will again need to be considered when applying our framework in this scenario.
\end{remark}

When kernel regression is used, a choice needs to be made regarding what kernel function and bandwidth should be used.  Generally, the choice of kernel function is not as crucial as the selection of an appropriate bandwidth (see \citealt{wand.jones_1995}, for example).  A popular choice of kernel function is the Gaussian kernel, while bandwidth selection may be undertaken in a variety of ways, for example using plug-in, bootstrap or cross-validation approaches (\citealt{kohler.etal_2014a}).

In this paper, we utilise an approach for selecting an optimal kernel bandwidth which is based on an idea first proposed by \citet{morikawa.etal_2017}.  The approach was described by those authors as a cross-validation approach.  The known data $(\boldsymbol{x}^*_i, \boldsymbol{x}_i, \boldsymbol{z}_i)$ taken from $A \cap B$ is used as training data to fit kernel regressions for the quantities
\begin{align}
    \begin{bmatrix}
    &\pi'(\boldsymbol{x}_i, \boldsymbol{z}_i; \hat{\boldsymbol{\theta}}) (1 - \pi(\boldsymbol{x}_i, \boldsymbol{z}_i;\hat{\boldsymbol{\theta}}))^{-1}O(\hat{\boldsymbol{\theta}};\boldsymbol{x}_i,\boldsymbol{z}_i) \nonumber \\
    &O(\hat{\boldsymbol{\theta}};\boldsymbol{x}_i,\boldsymbol{z}_i) \nonumber
\end{bmatrix}
\end{align}
using a set of candidate bandwidths $h$.  The kernel bandwidth which provides the best, in some sense, fit for these quantities over the data $A \cap B$ is chosen to be used in the full set of data to solve the score equation.

To enunciate the proposed bandwidth selection approach, first define
\begin{align}
    \hat{C}(\hat{\boldsymbol{\theta}}_h; \boldsymbol{z}_i) &= (\tilde{n}h^d)^{-1}\sum_{j \in A \cap B} \pi'(\boldsymbol{x}_j, \boldsymbol{z}_i; \hat{\boldsymbol{\theta}}_h) (1 - \pi(\boldsymbol{x}_j, \boldsymbol{z}_i;\hat{\boldsymbol{\theta}}_h))^{-1}O(\hat{\boldsymbol{\theta}}_h;\boldsymbol{x}_j,\boldsymbol{z}_i)K_h(\tilde{\boldsymbol{x}}_j, \tilde{\boldsymbol{x}}_i) \label{eqn5:c.hat} \\
    \hat{D}(\hat{\boldsymbol{\theta}}_h; \boldsymbol{z}_i) &= (\tilde{n}h^d)^{-1}\sum_{k \in A \cap B} O(\hat{\boldsymbol{\theta}}_h;\boldsymbol{x}_k, \boldsymbol{z}_i)K_h(\tilde{\boldsymbol{x}}_k, \tilde{\boldsymbol{x}}_i) \label{eqn5:d.hat}
\end{align}
for the overlapping sample $A \cap B$, where $\tilde{n}$ is the sample size feeding into the kernel estimation, $h$ is a candidate bandwidth and $d$ is the dimension of $\tilde{\boldsymbol{x}}$.  When $A$ comprises the full population then $A \cap B$ is equivalent to $B$.

From (\ref{eqn5:c.hat}) and (\ref{eqn5:d.hat}), and using (\ref{eqn5:kernel.exp}), one can see that
\begin{align}
    \frac{\hat{C}(\hat{\boldsymbol{\theta}}_h; \boldsymbol{z}_i)}{\hat{f}_1(\tilde{\boldsymbol{x}}_i)} &= E_1 \Bigl[\pi'(\boldsymbol{x}_i, \boldsymbol{z}_i; \hat{\boldsymbol{\theta}}_h) (1 - \pi(\boldsymbol{x}_i, \boldsymbol{z}_i;\hat{\boldsymbol{\theta}}_h))^{-1}O(\hat{\boldsymbol{\theta}}_h;\boldsymbol{x}_i,\boldsymbol{z}_i) | \tilde{\boldsymbol{x}}_i\Bigr] \text{ and} \nonumber \\
    \frac{\hat{D}(\hat{\boldsymbol{\theta}}_h; \boldsymbol{z}_i)}{\hat{f}_1(\tilde{\boldsymbol{x}}_i)} &=E_1\Bigl[ O(\hat{\boldsymbol{\theta}}_h;\boldsymbol{x}_i,\boldsymbol{z}_i | \tilde{\boldsymbol{x}}_i) \Bigr] \nonumber
\end{align}
where the expectations are kernel regression estimates and $\hat{f}_1(\boldsymbol{x}) = \sum_{i \in A \cap B}K_{h'}(\boldsymbol{x}_i - \boldsymbol{x})$ for a given bandwidth $h'$.  The bandwidth $h'$ may be chosen using established approaches (see for example \citealt{wand.jones_1995}).

The bandwidth selection approach seeks to minimise the overall error in these expectations.  Specifically, we choose $h$ out of a set of candidate bandwidths $H$ in order to minimise
\begin{align}
    K(h) = \sum_{i \in A} \delta_i \left\lVert
    \begin{bmatrix}
        \hat{C}(\hat{\boldsymbol{\theta}}_h; \boldsymbol{z}_i)/\hat{f}_1(\tilde{\boldsymbol{x}}_i) - \pi'(\boldsymbol{x}_i, \boldsymbol{z}_i; \hat{\boldsymbol{\theta}}_h) (1 - \pi(\boldsymbol{x}_i, \boldsymbol{z}_i;\hat{\boldsymbol{\theta}}_h))^{-1}O(\hat{\boldsymbol{\theta}}_h;\boldsymbol{x}_i,\boldsymbol{z}_i) \nonumber \\
        \hat{D}(\hat{\boldsymbol{\theta}}_h; \boldsymbol{z}_i)/\hat{f}_1(\tilde{\boldsymbol{x}}_i) - O(\hat{\boldsymbol{\theta}}_h;\boldsymbol{x}_i,\boldsymbol{z}_i)
    \end{bmatrix}
    \right\rVert^2 
\end{align} 

where $\left\lVert \cdot \right\rVert$ is the Euclidean norm and $\hat{\boldsymbol{\theta}}_h$ are the estimated participation probability model parameters estimated using bandwith $h$.

\section{Consistency and Variance Estimation} \label{sec5:consistency.variance}

In this section, we provide consistency properties of the IPW estimator formed using $\hat{\pi}^B_i$ estimated from the kernel score functions $\hat{s}(\boldsymbol{\theta})$ from (\ref{eqn5:score_kr}) or $\hat{s}_p(\boldsymbol{\theta})$ from (\ref{eqn5:score_kr_sample}).  We then provide details of a bootstrap approach to estimate the variance of the IPW estimator when the $\hat{\pi}^B_i$ are obtained using $\hat{\boldsymbol{\theta}}$ estimated from the kernel score function.

\begin{lemma} \label{lemma:kernel.popn.consistency}
    Under regularity conditions \textbf{C1} to \textbf{C4} in the Supplementary Materials, the kernel score function $\hat{s}(\boldsymbol{\theta})$ converges in probability to the observed score function $s(\boldsymbol{\theta})$ as $n_{AB} \to \infty$, where $n_{AB}$ is the size of the set $A \cap B$.
\end{lemma}

The proof of Lemma \ref{lemma:kernel.popn.consistency} is provided in the Supplementary Materials.

To examine consistency of the pseudo-likelihood $\hat{s}_p(\boldsymbol{\theta})$, we adopt an asymptotic framework used by other authors (see for example \citealt{chen.etal_2020}, \citealt{wang.etal_2021} and \citealt{scholtus.etal_2024a}).  Within this framework, suppose there is a sequence of finite populations $\mathcal{U}_t$ of size $N_t$ indexed by $t$.  Associated with each $\mathcal{U}_t$ are a probability sample $A_t$ of size $n_{A,t}$ and non-probability sample $B_t$ of size $N_{B,t}$.  The population size $N_t \to \infty$, while the sample sizes $n_{A,t} \to \infty$ and $n_{B,t} \to \infty$ as $t \to \infty$.  In what follows, the index $t$ is omitted, and the limiting process is represented by $N \to \infty$.

\begin{lemma} \label{lemma:kernel.samp.consistency}
    Under regularity conditions \textbf{C1} to \textbf{C7} in the Supplementary Materials, the kernel score function $\hat{s}_p(\boldsymbol{\theta})$ converges in probability to the observed score function $s(\boldsymbol{\theta})$ as $n_{AB} \to \infty$.
\end{lemma}

\begin{proof}
    To consider convergence of $\left\lVert \hat{s}_p(\boldsymbol{\theta}) - s(\boldsymbol{\theta}) \right\rVert$, we use the triangle inequality:
    \begin{align}
        \lVert \hat{s}_p(\boldsymbol{\theta}) - s(\boldsymbol{\theta}) \rVert &= \lVert \Bigl(\hat{s}_p(\boldsymbol{\theta}) - \hat{s}(\boldsymbol{\theta}) \Bigr)- \Bigl( \hat{s}(\boldsymbol{\theta}) - s(\boldsymbol{\theta}) \Bigr)\rVert \nonumber \\
        &\le \lVert \Bigl(\hat{s}_p(\boldsymbol{\theta}) - \hat{s}(\boldsymbol{\theta}) \Bigr) \rVert + \lVert \Bigl( \hat{s}(\boldsymbol{\theta}) - s(\boldsymbol{\theta}) \Bigr)\rVert \nonumber \\
        &\le o_p(n^{-1/4}) \nonumber
    \end{align}
    since by condition \textbf{C7} the first term on the right hand side of the inequality is $o_p(n^{-1/2})$, and by Lemma \ref{lemma:kernel.popn.consistency} the second term on the right hand side has been shown to be at least $o_p(n^{-1/4})$.
\end{proof}

Lemmas \ref{lemma:kernel.popn.consistency} and \ref{lemma:kernel.samp.consistency} have shown that $\hat{s}(\boldsymbol{\theta})$ and $\hat{s}_p(\boldsymbol{\theta})$ converge in probability to $s(\boldsymbol{\theta})$.  We next present Theorem \ref{thm:est.consistency} which looks at convergence of $\hat{\mu}_y$ under the observed score function $s(\boldsymbol{\theta})$.

\begin{theorem} \label{thm:est.consistency}
    Given assumptions \textbf{A1} to \textbf{A3}, and regularity conditions allowing the exchange of differentiability and integrability, hold.  When $\hat{\pi}^B_i$ is estimated by solving $s(\boldsymbol{\theta})=0$, then $E(\hat{\mu}_{IPW}) \to \mu_y$ and $\lvert \hat{\mu}_{IPW} - \mu_y \rvert \overset{p} \to 0$ as the sample sizes tend to infinity.
\end{theorem}

A proof is provided in the Supplementary Materials.

\subsection{Pseudo-population bootstrap for variance estimation} \label{sec5:bootstrap.variance}

To estimate the variance of $\hat{\mu}_y$, a pseudo-population bootstrap procedure is proposed based on the idea of \citet{booth.etal_1994}; see also \citet{mashreghi.etal_2016} for a review of bootstrap methods for finite populations.  The advantage of the pseudo-population bootstrap approach is that it allows one to conveniently take into account the overlapping sample in the calculation of bootstrap estimates.  \citet{liu.etal_2022} employed a similar bootstrap approach to take into account the possible dependency between the non-probability and reference samples.

The bootstrap procedure described in this subsection involves creating a pseudo-population using the data from $B$.  This means that data about $(y, \boldsymbol{x}^*, \boldsymbol{x}, \boldsymbol{z})$ is needed for all $i \in B$.  When $A$ comprises the full population this requirement will be fulfilled since values for $\boldsymbol{x}^*$ will be available for all units in the population.  Similarly, when $B$ is subject to SNAR missingness it will contain all the required data items since in this scenario $\boldsymbol{x}^*$ omits the variable of interest $y$ (recall Remark \ref{rem5:snar.framework}).

In the case when there is measurement error in the auxiliary variables in $A$ and $A$ is a sample from the population, $\boldsymbol{x}^*$ is only available for the units in $B \cap A$.  In this paper, we propose to produce imputed values $\hat{\boldsymbol{x}}^*_i$ for $i \in B \setminus A$ to fill the gap in the data for the pseudo-population bootstrap.  The imputed values may be obtained by fitting a model such as 
\begin{equation} \label{eqn: me defn}
    x^*_i = \beta_0 + \beta_1x_i + e_i
\end{equation}
using data in $A \cap B$, and then applying it to the records in $B$ to obtain $\hat{\boldsymbol{x}}^*_i, i \in B \setminus A$.  The augmented $B$ dataset, which now includes $\hat{\boldsymbol{x}}^*_i$ for $i \in B \setminus A$ and $\boldsymbol{x}^*_i$ for $i \in B \cap A$, is used to form the pseudo-population.

The pseudo-population bootstrap proceeds as follows.  Once any required imputation to obtain $\hat{\boldsymbol{x}}^*$ is undertaken, a pseudo-population of size $N$ is produced using the units in $B$.  Like \citet{liu.etal_2022}, the $\hat{\pi}^B_i$ are used as the basis for expanding the sample $B$ to form the population.  As a first step, the weights $1/\hat{\pi}^B_i$ are adjusted to ensure they sum to $N$, creating $\tilde{w}_i = \text{max}(1, N\hat{\pi}^B_i/\sum_{i \in B}(\hat{\pi}^B_i)^{-1})$.  The fixed part of the pseudo-population $\mathcal{U}^f$ is then created by replicating each unit $i$ of the original $B$ sample $k = \lfloor \tilde{w}_i \rfloor$ times.  The pseudo-population is completed by drawing $\mathcal{U}_c$ from $B$ using Poisson sampling with probability $\tilde{w}_i - \lfloor \tilde{w}_i \rfloor$ for unit $i$.  The final pseudo-population from which bootstrap samples are taken is $\mathcal{U}^* = \mathcal{U}_c \cup \mathcal{U}_f$.

\begin{algorithm}[ht]
  \caption{Bootstrap variance estimation for proposed IPW estimator}\label{alg5:me.bootvar}
  \begin{algorithmic}[1]
      \State If required, produce $\hat{\boldsymbol{x}}^*_i, i \in B \setminus A$ using a model fitted using the records in $A \cap B$
      \State Produce $\tilde{w}_i = \text{max}(1, N\hat{\pi}^B_i/\sum_{i \in B}(\hat{\pi}^B_i)^{-1})$
      \State Create the fixed part of the pseudo-population $\mathcal{U}^f$ by replicating each unit $i$ of the original $B$ sample $k = \lfloor \tilde{w}_i \rfloor$ times
      \State Create $\mathcal{U}^c$ by drawing a sample from $B$ using Poisson sampling with probability $\tilde{w}_i - \lfloor \tilde{w}_i \rfloor$ for unit $i$.  The final pseudo-population is given by combining $\mathcal{U}^f$ and $\mathcal{U}^c$ to form $\mathcal{U}^*$
      \While{$r \leq R$}
          \State Select a sample $S_A$ from $\mathcal{U}^*$ with size $n_A$, using the original sample design of $A$
          \State Select a sample $S_B$ from $\mathcal{U}^*$ with size $n_B$ 
          \State Calculate $\hat{\boldsymbol{\theta}}^r$ using $S_A$ and $S_B$ using the method outlined in Section \ref{sec5:method}
          \State Calculate $\hat{\pi}^{Br}_i$ for each unit $i \in S_B$ using $\hat{\boldsymbol{\theta}}^r$
          \State Calculate $\hat{\mu}^{(r)}_y$ using $\hat{\pi}^{Br}_i$
      \EndWhile
      \State Calculate the bootstrap variance using (\ref{eqn5:bootvar})
  \end{algorithmic}
\end{algorithm}

For each bootstrap iteration, bootstrap non-probability samples $S_B$ may be drawn from $\mathcal{U}^*$ using Poisson sampling with inclusion probability $\hat{\pi}^B_i$.  An alternative sampling scheme such as a fixed-size PPS sampling scheme may be used based on knowledge about the selection mechanism into $B$.  The bootstrap probability samples $S_A$ are drawn from $\mathcal{U}^*$ based on the sample design of the probability sample, with probabilities of selection $\pi^A_i$.  The samples $S_A$ and $S_B$ are combined using the proposed method to obtain $\hat{\pi}^{Br}_i$ and hence an estimate of the population quantity of interest, for example population means $\hat{\mu}^{(r)}_y$.  The bootstrap sampling procedure is repeated $R$ times to obtain $(\hat{\mu}^{(1)}_y, ..., \hat{\mu}^{(R)}_y)$.  The bootstrap variance is obtained through
\begin{align}
    \hat{Var}(\hat{\mu}_y) = \frac{1}{R - 1}\sum_{r=1}^{R}\Bigl( \hat{\mu}^{(r)}_y - \mu^*_y \Bigr)^2 \label{eqn5:bootvar}
\end{align}
where $\mu^*_y = N^{-1}\sum_{i \in \mathcal{U}^*}y_i$ is the mean of the $y$ variable in the pseudo-population (see \citealt{mashreghi.etal_2016}).

Algorithm \ref{alg5:me.bootvar} outlines the bootstrap variance procedure.

\section{Simulation Study} \label{sec5:simstudy}

A simulation study was conducted to assess the performance of the proposed approach and its bootstrap variance estimator.  A finite population of size $N = 20,000$ was constructed comprising a data item of interest $y$ and auxiliary variables $\boldsymbol{x} = (x_1, x_2, x_3, x_4)$.  The variable $y$ was related to $\boldsymbol{x}$ by the model
\begin{equation}
    y_i = f(\boldsymbol{x},\boldsymbol{\beta}) = 2 + 1.2x_{1i} + x_{2i} + x_{3i} + x_{4i} + \sigma\epsilon_i \nonumber
\end{equation}
with $x_{1i} = z_{1i}, x_{2i} = z_{2i} + 0.3x_{1i}, x_{3i} = z_{3i} + 0.2(x_{1i} + x_{2i}), x_{4i} = z_{4i} + 0.1(x_{1i} + x_{2i} + x_{3i})$, and $z_{1i} \sim \text{Bernoulli}(0.5), z_{2i} \sim \text{Uniform}(0,2), z_{3i} \sim \text{Exponential}(1)$, and $z_{4i} \sim \chi^2(4)$.  The error terms $\epsilon_i$ were independent and identically distributed as $\text{Normal}(0,1)$.  For these simulations, $\sigma$ was set such that the correlation coefficient $\rho$ between $y$ and $\boldsymbol{x}^T\boldsymbol{\beta}$ was 0.3.

In this simulation the data item $x_{4}$ may be subject to measurement error.  The performance of our proposed method was examined under four slightly different measurement error models:
\begin{enumerate}
    \item $x^*_4 = x_4 + \gamma$, where $\gamma = \text{Normal}(0,1)$
    \item $x^*_4 = 0.8x_4 + \gamma$, where $\gamma = \text{Normal}(0,1)$
    \item $x^*_4 = x_4 + \gamma$, where $\gamma = \text{Normal}(0,2)$
    \item $x^*_4 = 10x_4^{1/4} + \gamma$, where $\gamma = \text{Normal}(0,0.7)$
\end{enumerate}

Case 1 relates to the classical measurement error scenario where the mis-measured variable is given by the true value of the variable plus some random error.  In Case 2, the mis-measured variable tends to have a lower value compared with the true value of the variable.  Case 3 is similar to Case 1, but the error term has a larger standard deviation.  Finally, Case 4 represents a non-linear measurement error model.

The true participation probabilities into the non-probability sample, $\pi^B_i$, were given by
\begin{equation}
    \text{log}\Bigl(\frac{\pi^B_i}{1 - \pi^B_i} \Bigr) = \theta_0 + \theta_1x_{1i} + \theta_2x_{2i} + \theta_3x_{3i} + \theta_4x_{4i} \nonumber
\end{equation}
where the intercept $\theta_0$ was chosen to ensure the target sample size for $N_B$ is achieved, and $(\theta_1, \theta_2, \theta_3, \theta_4) = (0.1, 0.2, 0.3, 0.4)$.  Poisson sampling was used to select the sample $B$.  Pareto probability proportional to size (PPS) sampling was used to select the probability sample $A$.  In general, the inclusion probabilities for $\pi^A_i$ were proportional to $z_i = c + x_{3i}$, where $c$ was chosen to ensure the ratio max($z_i$)/min($z_i$) = 20.

For each of the measurement error scenarios, $R = 2,000$ simulation iterations were executed, with $B=50$ bootstrap iterations used to calculate a bootstrap variance using the approach outlined in Algorithm \ref{alg5:me.bootvar}.  Since in this simulation study the samples $A$ do not comprise the full population, imputed values $\hat{x}^*_{4,i}, i \in B \setminus A$ were produced for the bootstrap pseudo-population using a linear model fitted on the common sample $A \cap B$ in each simulation iteration:
\begin{align}
    \hat{x}^*_{4,i} = \hat{\beta}_0 + \hat{\beta}_1 x_{4,i} + \hat{\gamma} \label{eqn5:x4.impute}
\end{align}
where $\hat{\beta}_0$, $\hat{\beta}_1$ were the fitted model parameters and $\hat{\gamma}$ was the estimated residual error term.

A Gaussian kernel was used to implement the kernel smoothing for the proposed approach.  An optimal kernel bandwidth for $\hat{C}$ and $\hat{D}$ was chosen using the bandwidth selection method described in Section \ref{sec5:method}.  The candidate bandwidths used were of a form based on the data-based rule in \citet{scott_2014}:
\begin{align}
    h(p) = n^{-p}\hat{\sigma}_x \label{eqn5:bw.form}
\end{align}
where $n$ was the sample size feeding in to the kernel estimation, $\hat{\sigma}_x$ was the estimated standard deviation of the variable $x$, and $p$ was allowed to range from 0.1 to 0.5.  To estimate the densities $\hat{f}_1(\tilde{\boldsymbol{x}})$ used in the bandwidth selection approach, kernel smoothing was again utilised, with a Gaussian kernel and bandwidths $n^{-0.4}\hat{\sigma}_x$.

For the simulation exercise, the performance of the proposed approach was assessed for values of $N_B$ of 2,000 and 8,000.  The reference sample size $n_A$ was kept at a constant 1,000.  Results were produced for the following estimators:
\begin{itemize}
    \item Naive - Simple sample mean from $B$
    \item OE - The IPW estimator which uses participation probabilities derived using the approach of \citet{ang.etal_2024a} (see Remark \ref{rem:oe.likelihood}), without any correction for measurement error
    \item OE\_Model - The OE IPW estimator with measurement error corrected by fitting a model $x^*_4 = \hat{\beta}_0 + \hat{\beta}_1x_4$ using the common sample $A \cap B$, and then applying it to the sample $A$ to estimate $\hat{x}_4$
    \item KW - The IPW estimator which uses participation probabilities estimated using the data integration approach proposed in \citet{kim.wang_2019}
    \item KW\_Model - The KW approach, with measurement error now corrected by fitting a model $x^*_4 = \hat{\beta}_0 + \hat{\beta}_1x_4$ using the common sample $A \cap B$, and then applying it to the sample $A$ to estimate $\hat{x}_4$
    \item ME - The proposed method for addressing measurement error in some auxiliary variables in $A$
    \item OE\_True - The OE IPW estimator assuming there is no measurement error for $x_4$ in $A$
\end{itemize}

The seven estimators were evaluated using the Monte Carlo Relative Bias (RB) and Relative Root Mean Squared Error (RRMSE).

Table \ref{tab5:MEsim_results1a} compares the RB and RRMSE of the estimators for the classical measurement error case, Case 1.  The OE\_True results represent a benchmark performance under an ideal case where there is no covariate measurement error, and they are unbiased with the lowest RRMSE.  The OE, KW, OE\_Model and KW\_Model estimates have a small bias as a result of bias in the participation probability model parameters.  On the other hand, the resulting estimates of mean from the proposed ME approach are unbiased and have comparable RRMSE values to the true model.

% Table generated by Excel2LaTeX from sheet 'SimResults_ME_Sample', Seed 202506
\begin{table}[ht]
    \centering
    \caption{Comparison of estimators under measurement error - Case 1}
      \begin{tabular}{lrrrr}
        \hline
      \multicolumn{1}{c}{} & \multicolumn{2}{c}{$N_B=2,000$} & \multicolumn{2}{c}{$N_B=8,000$} \\
       & \%RB & \%RRMSE & \%RB & \%RRMSE \\
        \hline
            Naive       & 42.67 & 42.73 & 19.30 & 19.32 \\
            OE          & 1.55  & 4.53  & 1.08  & 1.84 \\
            OE\_Model   & 1.56  & 4.57  & 1.09  & 1.85 \\
            KW          & 1.17  & 5.57  & 0.94  & 2.19 \\
            KW\_Model   & 1.18  & 5.61  & 0.95  & 2.20 \\
            ME          & -0.13 & 4.57  & 0.09  & 1.59 \\
            OE\_True    & 0.00  & 4.40  & -0.01 & 1.53 \\
      \hline
      \end{tabular}
    \label{tab5:MEsim_results1a}
\end{table}

In this measurement error scenario, creating a model to correct measurement error does not help at all, as evidenced by the KW\_Model and OE\_Model results.  To see why this is the case, recall that in the classical measurement error scenario the mis-measured value of a variable is equal to the true value plus some random error.  The fitted model would thus result in predicted true values that are approximately equal to the original mis-measured values.  This is why the KW\_Model and OE\_Model results are very similar to the KW and OE results respectively.  Note that it may seem reasonable to instead employ a measurement error model which includes a fitted error term, for example $x^*_4 = \hat{\beta}_0 + \hat{\beta}_1x_4 + \hat{\sigma}$ where $\hat{\sigma}$ is estimated from the model fitting process.  However, in the simulation study this approach was found to actually lead to worse results.

Table \ref{tab5:MEsim_results1b} compares the RB and RRMSE of the various estimators under measurement error for Case 2.  The OE and KW estimates are again biased and have a fairly large RRMSE because they have not corrected for measurement error.  Fitting a measurement error model, imputing a corrected $\hat{x}_4$ value, and using that in the participation probability model is able to correct much, but not all of the measurement error bias.  The RRMSE results for the KW\_Model and OE\_Model approaches are better than if no action had been taken at all.  The proposed ME approach to correct measurement error is shown to be effective.  The estimates produced by the approach are unbiased.  In particular, the RRMSE results for the proposed ME approach are only slightly higher than the benchmark OE\_True results.

% Table generated by Excel2LaTeX from sheet 'SimResults_ME_Sample', Seed 202506
\begin{table}[ht]
    \centering
    \caption{Comparison of estimators under measurement error - Case 2}
      \begin{tabular}{lrrrr}
        \hline
      \multicolumn{1}{c}{} & \multicolumn{2}{c}{$N_B=2,000$} & \multicolumn{2}{c}{$N_B=8,000$} \\
       & \%RB & \%RRMSE & \%RB & \%RRMSE \\
        \hline
            Naive       & 42.67 & 42.73 & 19.30 & 19.32 \\
            OE          & -7.18 & 8.92  & -2.97 & 3.41 \\
            OE\_Model   & 2.39  & 4.88  & 1.63  & 2.20 \\
            KW          & -7.67 & 10.05 & -3.18 & 3.92 \\
            KW\_Model   & 2.02  & 5.79  & 1.50  & 2.45 \\
            ME          & 0.10  & 4.64  & 0.23  & 1.63 \\
            OE\_True    & 0.00  & 4.40  & -0.01 & 1.53 \\
      \hline
      \end{tabular}
    \label{tab5:MEsim_results1b}
\end{table}

Table \ref{tab5:MEsim_results1c} provides the RB and RRMSE values of the estimators under Case 3.  In this case the mis-measured variable is equal to the true value plus an error that has a larger standard deviation than Case 1.  From these results, the OE, OE\_Model, KW, and KW\_Model approaches again produce biased estimates with fairly large RRMSE values.  Like Case 1, fitting a model to correct the measurement error does not help with reducing the bias since the predicted true values are approximately equal to the original mis-measured values.

By including the measurement error correction as part of the likelihood maximisation process, the ME approach mitigates the bias, and it also has a reduced RRMSE compared to the OE and OE\_Model approaches.  However, it does not completely remove the bias.  On further inspection, the inability to remove all bias seems to stem from the small overlapping sample between $A$ and $B$.  When $N_B=2,000$ and $n_A=1,000$, for example, the overlapping sample is around 120.  When we used $N_B=2,000$ and $n_A=4,000$ the overlapping sample increased to about 500 units, and the RB for the ME approach dropped to 1.7.  When there is a lot of noise in the measurement error, the ME approach requires more overlapping sample to achieve an accurate result.

% Table generated by Excel2LaTeX from sheet 'SimResults_ME_Sample', Seed 202506
\begin{table}[ht]
    \centering
    \caption{Comparison of estimators under measurement error - Case 3}
      \begin{tabular}{lrrrr}
        \hline
      \multicolumn{1}{c}{} & \multicolumn{2}{c}{$N_B=2,000$} & \multicolumn{2}{c}{$N_B=8,000$} \\
       & \%RB & \%RRMSE & \%RB & \%RRMSE \\
        \hline
            Naive       & 42.67 & 42.73 & 19.30 & 19.32 \\
            OE          & 5.66  & 6.88  & 3.49  & 3.76 \\
            OE\_Model   & 5.71  & 7.02  & 3.51  & 3.79 \\
            KW          & 5.32  & 7.37  & 3.38  & 3.82 \\
            KW\_Model   & 5.37  & 7.50  & 3.40  & 3.85 \\
            ME          & 2.49  & 5.31  & 1.41  & 2.21 \\
            OE\_True    & 0.00  & 4.40  & -0.01 & 1.53 \\
      \hline
      \end{tabular}
    \label{tab5:MEsim_results1c}
\end{table}

Finally, Table \ref{tab5:MEsim_results1d} shows the RB and RRMSE values of the estimators under Case 4.  In this scenario, the OE approach severely underestimates the participation probabilities, leading to very large RB and RRMSE results.  The RB and RRMSE results for the KW approach are also very poor in this sccenario.  Fitting a linear measurement error model to the common units and using the predicted true values of $x_4$ reduces the bias and RRMSE significantly, but does not eliminate it.  In contrast, the proposed ME approach is unbiased and has RRMSE values very close to the gold standard OE\_True model.

% Table generated by Excel2LaTeX from sheet 'SimResults_ME_Sample', Seed 202506
\begin{table}[ht]
    \centering
    \caption{Comparison of estimators under measurement error - Case 4}
      \begin{tabular}{lrrrr}
        \hline
      \multicolumn{1}{c}{} & \multicolumn{2}{c}{$N_B=2,000$} & \multicolumn{2}{c}{$N_B=8,000$} \\
       & \%RB & \%RRMSE & \%RB & \%RRMSE \\
        \hline
            Naive       & 42.67 & 42.73 & 19.30 & 19.32 \\
            OE          & 243.33 & 243.42 & 263.83 & 263.88 \\
            OE\_Model   & 6.64  & 7.71  & 3.28  & 3.55 \\
            KW          & 243.34 & 243.94 & 263.78 & 263.99 \\
            KW\_Model   & 6.27  & 8.01  & 3.15  & 3.61 \\
            ME          & 0.08  & 4.66  & 0.03  & 1.59 \\
            OE\_True    & 0.00  & 4.40  & -0.01 & 1.53 \\
      \hline
      \end{tabular}
    \label{tab5:MEsim_results1d}
\end{table}

The performance of the bootstrap variance estimator described in Section \ref{sec5:consistency.variance} for the ME approach is assessed in Table \ref{tab5:MEsim_results2} for each of the four scenarios.  The table contains the coverage probabilities (CP) and average length (AL) of the 95\% confidence intervals for the estimates of population mean for two implementations of the ME bootstrap.  The first implementation assumes $x^*_4$ is available in the full population, and is used as a benchmark to assess the second implementation which assumes $x^*_4$ is only available in $A$.

% Table generated by Excel2LaTeX from sheet 'SimResults_ME_Sample', Seed 202506
\begin{table}[ht]
    \centering
    \caption{Performance of variance estimators under measurement error}
      \begin{tabular}{llrrrr}
        \hline
      \multicolumn{2}{c}{} & \multicolumn{2}{c}{$N_B=2,000$} & \multicolumn{2}{c}{$N_B=8,000$} \\
       & & CP & AL & CP & AL \\
        \hline
        \multicolumn{6}{l}{Case 1} \\
           & ME Bootstrap ($x^*_4$ in population)    & 0.954  & 1.757  & 0.950  & 0.599 \\
           & ME Bootstrap ($x^*_4$ in $A$ only)    & 0.957  & 1.764  & 0.949  & 0.599 \\
        \hline
        \multicolumn{6}{l}{Case 2} \\
           & ME Bootstrap ($x^*_4$ in population)   & 0.950  & 1.764  & 0.946  & 0.610 \\
            & ME Bootstrap ($x^*_4$ in $A$ only)   & 0.954  & 1.776  & 0.948  & 0.611 \\
        \hline
        \multicolumn{6}{l}{Case 3} \\
           & ME Bootstrap ($x^*_4$ in population)   & 0.918  & 1.873  & 0.922  & 0.766 \\
           & ME Bootstrap ($x^*_4$ in $A$ only)   & 0.917  & 1.881  & 0.920  & 0.765 \\
        \hline
        \multicolumn{6}{l}{Case 4} \\
           & ME Bootstrap ($x^*_4$ in population)   & 0.951  & 1.765  & 0.950  & 0.596 \\
           & ME Bootstrap ($x^*_4$ in $A$ only)   & 0.959  & 1.811  & 0.961  & 0.622 \\
      \hline
      \end{tabular}
      \begin{minipage}{12cm}
		\vspace{0.1cm}
        \spacingset{1.0}
		\small Note: CP are the estimated Coverage Probabilities, while AL denotes the average length of the 95\% confidence intervals
	   \end{minipage}
    \label{tab5:MEsim_results2}
\end{table}

In all scenarios, the CP and AL values for the ME bootstrap when $x^*_4$ is only available in $A$ are very close to the CP and AL values for the ME boostrap when $x^*_4$ is available for the full population.  The imputation used for $x^*_4$ is effective, especially for Case 1, Case 2 and Case 3 where a linear relationship between $x_4$ and $x^*_4$ exists.

In Case 4, where there is a non-linear relationship between $x_4$ and $x^*_4$, the imputation leads to a slight over-estimation of the bootstrap variance, resulting in slightly higher CP and AL values.  This is understandable since from (\ref{eqn5:x4.impute}) the imputation of $x^*_4$ assumes a linear relationship with $x_4$.  The bootstrap variance results could be improved, for example by employing an imputation model more suited to the non-linear measurement error.  In the discussion that follows, the focus will be on the ME bootstrap results where $x^*_4$ is assumed to be available only on $A$.

For Case 1 and Case 2, the CP values for the ME bootstrap variance estimator under both sample sizes for $N_B$ are very close to the nominal value of 0.95.  For Case 3, the CP of the bootstrap estimator is still around 0.92; it is slightly lower than the nominal value of 0.95 due to the residual bias in the estimates.  In Case 4, as has been noted, the CP values are slightly higher than the nominal value given that the imputation of $x^*_4$ assumes a different measurement error model compared with reality.

The ALs for the ME bootstrap variance estimator are all reasonable in magnitude relative to the population mean of around 9.4.

\section{Real-world data simulation} \label{sec5:real.data}

In this section, the performance of the proposed framework is examined using some real business data.  A population dataset was created Australian Business Activity Statement (BAS) data from the July-September 2022-2023 quarter.  The BAS data was sourced using the Business Longitudinal Analysis Data Environment (BLADE) standard module and accessed through the Australian Bureau of Statistics (ABS) DataLab.  Records designated as live businesses, in the private or public sector, and with an active Goods and Services Tax (GST) payer role, were extracted.  Pay-As-You-Go (PAYG) data relating to the number of employees in each business was also obtained for the 2022-2023 financial year.  Due to computational considerations, this empirical study was restricted to businesses in the Manufacturing industry division.  The resulting population consisted of 37,218 units.

Total sales excluding GST (turnover) was used as the variable of interest.  The auxiliary data items were total salary, wages and other payments (wages) and an employee count (employees) for the business.  A version of the wages data item containing measurement error was created by merging onto the population file the wages value for each business, if available, from the July-September 2020-2021 quarter BAS data.  This represented an out-of-date frame variable.  If a business did not have any wages data from that quarter, the 2022-2023 quarter BAS data was assigned to it.  In other words, when the 2020-2021 data was not available, the business was assumed to have no measurement error.  Since they were very right-skewed variables, the with- and without-measurement error wages variables were both transformed using a log transformation before being used.  See the Supplementary Materials for a binned scatterplot showing the relationship between the true wages and the error-prone wages variables.

The performance of our proposed ME approach for addressing covariate measurement error in $A$ was assessed through a simulation study using the data from this population.  For each iteration of the simulation, a non-probability dataset was selected using Poisson sampling.  Selection into the non-probability dataset was driven by a logistic model with log(wages+1) and employees as the covariates:
\begin{align}
    \text{log}\Bigl(\frac{\pi^B_i}{1 - \pi^B_i} \Bigr) = \theta_0 + 0.05x_{\text{employees}} + 0.1x_{\text{log(wages)}} \nonumber
\end{align}
where $\theta_0$ was chosen so that the non-probability sample size $N_B=3,000$.  The data items available on this dataset were turnover, employees and true wages.

For each simulation iteration, a reference probability sample of size $n_A=5,000$ was selected using PPS sampling, with the log of the measurement-error version of wages used as the measure of size.  For businesses where this variable had a value greater than 1.5, the value was set to 1.5.  The inclusion probabilities $\pi^A_i$ were proportional to $z_i = c + \text{log}(\text{wages}_{me,i})$, where $c$ was chosen to ensure the ratio $\text{max}(z_i)/\text{min}(z_i)$ was equal to 20 and $\text{wages}_{me,i}$ was the two-year-old wages value for unit $i$ exhibiting measurement-error.  For each simulation iteration, a probability sample of size $n_A=5,000$ was chosen using Pareto sampling.  The data items available on the reference samples were employees and the measurement-error version of wages.

For this simulation using the real-world data, $R=1,000$ iterations were run.  The performance of the following estimators was assessed:
\begin{itemize}
    \item Naive - Simple sample mean from $B$
    \item OE - The Overlap-Excluded (OE) IPW estimator from \citet{ang.etal_2024a} without any correction for measurement error
    \item ME - The proposed method for addressing measurement error in $A$, where kernel bandwidths were chosen using the bandwidth selection method described in Section \ref{sec5:method} using candidate bandwidths of the form (\ref{eqn5:bw.form}), and $\hat{f}_1(\tilde{\boldsymbol{x}})$ estimated using kernel smoothing with a Gaussian kernel and bandwidths $n^{-0.4}\hat{\sigma}_{x}$
    \item OE\_True - The OE IPW estimator when there is no measurement error for wages in $A$
\end{itemize}

% Table generated from DataLab
\begin{table}[ht]
    \centering
    \caption{Performance of proposed method using real-world data}
      \begin{tabular}{lrr}
        \hline
        & \%RB & \%RRMSE \\
        \hline
            Naive       & 505.57 & 505.63 \\
            OE          & -14.54 & 15.59 \\
            ME          & 1.44  & 7.00 \\
            OE\_True    & 0.02  & 4.20 \\
      \hline
      \end{tabular}
    \label{tab5:realdata_results}
\end{table}

Table \ref{tab5:realdata_results} shows the RB and RRMSE values for the estimators.  The Naive estimate is severely biased and has a very large RRMSE.  The OE estimator corrects for the selection bias, but because the wages variable in $A$ is prone to measurement error, the resulting estimate still shows a large bias and RRMSE.  The ME approach removes most of the bias, although there is still a small residual bias remaining.  Its RRMSE is also much smaller than the RRMSE for the OE method, although about 3 percentage points larger than the gold standard OE\_True estimator.

\section{Addressing error-prone auxiliary variables in $B$} \label{sec5:cov.error.B}

The approach proposed in this paper has focussed on the case where there is measurement error in some of the auxiliary variables present on the probability sample $A$.  However, measurement error may also be present for auxiliary variables on the non-probability sample $B$.  In this section, we briefly examine how to treat this case.  Again, to begin we assume that $A$ corresponds to a population-level dataset, and construct the observed likelihood function as
\begin{equation} \label{eqn5:obs.likelihood.covB}  
  L_b(\boldsymbol{\theta}) = \prod_{i \in \mathcal{U}}  \Bigl\{ \int\pi(\boldsymbol{x},\boldsymbol{z}_i;\boldsymbol{\theta}) f(\boldsymbol{x}|\boldsymbol{x}^*_i, \boldsymbol{z}_i) \text{d}\boldsymbol{x} \Bigr\}^{\delta_i} (1-\pi(\boldsymbol{x}_i,\boldsymbol{z}_i;\boldsymbol{\theta}))^{1-\delta_i}.
\end{equation}

The corresponding log-likelihood is
\begin{equation} \label{eqn5:obs.loglikelihood.covB}   
  l_b(\boldsymbol{\theta}) = \sum_{i \in \mathcal{U}} \delta_i \text{log} \Bigl\{ \int \pi(\boldsymbol{x}, \boldsymbol{z}_i; \boldsymbol{\theta})f(\boldsymbol{x}|\boldsymbol{x}^*_i, \boldsymbol{z}_i)\text{d}\boldsymbol{x} \Bigr\} + \sum_{i \in \mathcal{U}}(1 - \delta_i)\text{log}  (1 - \pi(\boldsymbol{x}_i, \boldsymbol{z}_i; \boldsymbol{\theta}))
\end{equation}

and, assuming it is possible to interchange differentiation and integration, the score function is given by
\begin{align} 
  s_b(\boldsymbol{\theta}) &= \frac{\partial l_b} {\partial \boldsymbol{\theta}} \nonumber \\ 
  &= \sum_{i \in \mathcal{U}} \delta_i \frac{\int \pi'(\boldsymbol{x}, \boldsymbol{z}_i; \boldsymbol{\theta})f(\boldsymbol{x} | \boldsymbol{x}^*_i, \boldsymbol{z}_i)\text{d}\boldsymbol{x}}{\int \pi(\boldsymbol{x}, \boldsymbol{z}_i; \boldsymbol{\theta}) f(\boldsymbol{x} | \boldsymbol{x}^*_i, \boldsymbol{z}_i)\text{d}\boldsymbol{x} } - \sum_{i \in \mathcal{U}}(1 - \delta_i)\frac{ \pi'(\boldsymbol{x}_i, \boldsymbol{z}_i; \boldsymbol{\theta})}{(1 - \pi(\boldsymbol{x}_i, \boldsymbol{z}_i; \boldsymbol{\theta}))}. \label{eqn5:score.covB}
\end{align}

The density $f(\boldsymbol{x} | \boldsymbol{x}^*, \boldsymbol{z})$ is not known, but the relationship (\ref{eqn5:distreln}) can be used to reformulate the score function as
\begin{align}
    s_b(\boldsymbol{\theta}) &= \sum_{i \in \mathcal{U}} \delta_i \frac{\int \pi'(\boldsymbol{x}, \boldsymbol{z}_i; \boldsymbol{\theta})f_1(\boldsymbol{x} | \boldsymbol{x}^*_i, \boldsymbol{z}_i){P(\delta_i=1|\boldsymbol{x}^*_i, \boldsymbol{z}_i)}/\pi(\boldsymbol{x}, \boldsymbol{z}_i;\boldsymbol{\theta}) \text{d}\boldsymbol{x}}{\int \pi(\boldsymbol{x}, \boldsymbol{z}_i; \boldsymbol{\theta}) f_1(\boldsymbol{x} | \boldsymbol{x}^*_i, \boldsymbol{z}_i){P(\delta_i=1|\boldsymbol{x}^*_i, \boldsymbol{z}_i)}/\pi(\boldsymbol{x}, \boldsymbol{z}_i;\boldsymbol{\theta}) \text{d}\boldsymbol{x} } \nonumber \\
    &- \sum_{i \in \mathcal{U}}(1 - \delta_i)\frac{ \pi'(\boldsymbol{x}_i, \boldsymbol{z}_i; \boldsymbol{\theta})}{(1 - \pi(\boldsymbol{x}_i, \boldsymbol{z}_i; \boldsymbol{\theta}))} \nonumber \\
    &= \sum_{i \in \mathcal{U}} \delta_i \frac{\int \pi'(\boldsymbol{x}, \boldsymbol{z}_i; \boldsymbol{\theta}) \pi(\boldsymbol{x}, \boldsymbol{z}_i;\boldsymbol{\theta})^{-1} f_1(\boldsymbol{x} | \boldsymbol{x}^*_i, \boldsymbol{z}_i) \text{d}\boldsymbol{x}}{\int f_1(\boldsymbol{x} | \boldsymbol{x}^*_i, \boldsymbol{z}_i) \text{d}\boldsymbol{x} } - \sum_{i \in \mathcal{U}}(1 - \delta_i)\frac{ \pi'(\boldsymbol{x}_i, \boldsymbol{z}_i; \boldsymbol{\theta})}{(1 - \pi(\boldsymbol{x}_i, \boldsymbol{z}_i; \boldsymbol{\theta}))} \nonumber \\ 
    &= \sum_{i \in \mathcal{U}} \delta_i \int \pi'(\boldsymbol{x}, \boldsymbol{z}_i; \boldsymbol{\theta}) \pi(\boldsymbol{x}, \boldsymbol{z}_i;\boldsymbol{\theta})^{-1} f_1(\boldsymbol{x} | \boldsymbol{x}^*_i, \boldsymbol{z}_i) \text{d}\boldsymbol{x} - \sum_{i \in \mathcal{U}}(1 - \delta_i)\frac{ \pi'(\boldsymbol{x}_i, \boldsymbol{z}_i; \boldsymbol{\theta})}{(1 - \pi(\boldsymbol{x}_i, \boldsymbol{z}_i; \boldsymbol{\theta}))}.\label{eqn5:score_subf1.covB}
\end{align}

Using the kernel regression approach to evaluate the integral, one can obtain
\begin{align}
    \hat{s}_b(\boldsymbol{\theta})
    &= \sum_{i \in \mathcal{U}} \delta_i \frac{ \sum_{j \in B} \pi'(\boldsymbol{x}_j, \boldsymbol{z}_i; \boldsymbol{\theta}) \pi(\boldsymbol{x}_j, \boldsymbol{z}_i;\boldsymbol{\theta})^{-1} K_h(\tilde{\boldsymbol{x}}_j, \tilde{\boldsymbol{x}}_i)}{\sum_{k \in B} K_h(\tilde{\boldsymbol{x}}_j, \tilde{\boldsymbol{x}}_i)} - \sum_{i \in \mathcal{U}}(1 - \delta_i)\frac{ \pi'(\boldsymbol{x}_i, \boldsymbol{z}_i; \boldsymbol{\theta})}{(1 - \pi(\boldsymbol{x}_i, \boldsymbol{z}_i; \boldsymbol{\theta}))} \nonumber \\ 
    &= \sum_{i \in \mathcal{U}} \delta_i \sum_{j \in B} \pi'(\boldsymbol{x}_j, \boldsymbol{z}_i; \boldsymbol{\theta}) \pi(\boldsymbol{x}_j, \boldsymbol{z}_i;\boldsymbol{\theta})^{-1}w_{b,ij} - \sum_{i \in \mathcal{U}}(1 - \delta_i)\frac{ \pi'(\boldsymbol{x}_i, \boldsymbol{z}_i; \boldsymbol{\theta})}{(1 - \pi(\boldsymbol{x}_i, \boldsymbol{z}_i; \boldsymbol{\theta}))}  \label{eqn5:scorekernel.covB}
\end{align}
where $w_{b,ij} = K_h(\tilde{\boldsymbol{x}}_j, \tilde{\boldsymbol{x}}_i) / \sum_{k \in B} K_h(\tilde{\boldsymbol{x}}_k, \tilde{\boldsymbol{x}}_i)$ and $\tilde{\boldsymbol{x}}=(\boldsymbol{x}^*,\boldsymbol{z})$ is a combined vector of covariates.

When $A$ is a subsample of the population, then one can substitute sample-weighted quantities into (\ref{eqn5:scorekernel.covB}) in place of their population counterparts to obtain
\begin{align}
    \hat{s}_{bp}(\boldsymbol{\theta})
    &= \sum_{i \in \mathcal{U}} \delta_i \sum_{j \in A \cap B} \pi'(\boldsymbol{x}_j, \boldsymbol{z}_i; \boldsymbol{\theta}) \pi(\boldsymbol{x}_j, \boldsymbol{z}_i;\boldsymbol{\theta})^{-1} \tilde{w}_{b,ij} - \sum_{i \in \mathcal{A}}d^A_i(1 - \delta_i)\frac{ \pi'(\boldsymbol{x}_i, \boldsymbol{z}_i; \boldsymbol{\theta})}{(1 - \pi(\boldsymbol{x}_i, \boldsymbol{z}_i; \boldsymbol{\theta}))}  \label{eqn5:scorekernel.covB.sample} \\
    &\text{where } \tilde{w}_{b,ij} = \frac{d^A_jK_h(\tilde{\boldsymbol{x}}_j, \tilde{\boldsymbol{x}}_i)}{\sum_{k \in B} d^A_kK_h(\tilde{\boldsymbol{x}}_k, \tilde{\boldsymbol{x}}_i)}. \nonumber
\end{align}

Examining the properties of the estimator that results from solving the score equation arising from (\ref{eqn5:scorekernel.covB.sample}), and assessing its performance empirically, are left for future work.  We note that one possible extension of the framework developed in this section is to help address selection bias in the case when the reference sample $A$ collects $(\boldsymbol{x},y)$ and $B$ collects $\boldsymbol{x}$ only.  Within the framework proposed in this section, the missing $y$ on $B$ can be thought of as an extreme form of measurement error, where $y$ is not measured at all.  Assessing the effectiveness of this approach to handle selection bias in this situation is also left for future work.

\section{Concluding Remarks} \label{sec5:conclusion}

The reference sample used to correct selection bias in a non-probability dataset may contain auxiliary (or covariate) information that is subject to measurement error.  Using the error-prone auxiliary variables to fit a participation probability model may lead to biased estimation of the model parameters.  In this paper a framework has been proposed to simultaneously correct for both measurement error in the reference sample and selection bias in the non-probability sample.  The framework is applicable in the case when it is possible to link the records in the reference and non-probability samples.  The proposed framework can also be applied in the context of SNAR missingness in the non-probability sample.

In both a simulation and real-world data setting, the proposed framework effectively mitigates the bias due to measurement error and selection bias, resulting in unbiased estimates of population mean when the measurement error is not as noisy, with RRMSE values very close to the ideal RRMSE achieved when there is no measurement error.  When there was greater variability in the measurement error, the proposed approach was still able to address most of the selection bias.

Two possible reasons for the small residual bias in some of the simulation results are the amount of overlap between the reference and non-probability samples, and the kernel bandwidth chosen.  Ensuring there is a reasonable overlap between the two samples is an important consideration for correcting measurement error.  How to design an efficient reference sample with sufficient overlap between it and the non-probability sample to correct for measurement error is a future research topic.  The bandwidth selection approach used in this chapter to choose the kernel bandwidth was effective in finding an optimal bandwidth for a given set of candidate bandwidths.  However, more effective bias reduction may potentially be obtained by widening the set of candidate bandwidths to choose from.  This is an area for future research.

The simulation and real-world data results show that not taking any action to resolve measurement error is not advisable, as the resulting estimates of population mean remain biased.  It was found that fitting a simple linear measurement error model with the common sample, and then using the model to impute corrected values for the affected variables, could be beneficial.  However, the resulting estimates retained some measure of bias, and did not perform as well as our proposed approach.

The proposed framework requires the ability to link the data in the reference and non-probability samples, and this may not be possible to do in reality, or to do it perfectly.  One possible extension of this work would be to explore how the approach may be applied when there is imperfect linkage between the two samples, or if linkage is not possible at all.

Finally, the proposed approach employs kernel regression to estimate the integrals in the score function.  In addition to the bandwidth selection challenges already mentioned, the kernel approach was found to be fairly computationally intensive as sample sizes increased.  As such, a possible extension for this work could be to explore alternative non-parametric approaches to estimate the integrals which are not affected by these issues.

%%%%%%%%%%%%%%%%%%%%%%%%%%%%%%%%%

\bigskip

\section*{Data availability}

The R code to produce the artificial population and conduct the comparison of estimators for the simulation study is at \url{https://github.com/skrawk/np-errorprone-aux}.

\bigskip

\section*{Acknowledgements, Disclaimer and Funding}

The views expressed in this paper are those of the authors and do not necessarily represent the views of the Australian Bureau of Statistics. Where quoted or used, they should be attributed clearly to the authors.  The first author was supported by funding from the Sir Roland Wilson Foundation and the Australian Bureau of Statistics.

\bigskip

\section*{Supplementary Material}

The supplementary material mentioned in this paper can be found on the publisher's website.  This additional information includes i) a derivation of the kernel score function for the special SNAR case, ii) conditions for consistency of the kernel score function, iii) a proof of consistency of the kernel score function for the error-prone auxiliary information case, iv) proof of consistency of the IPW estimator, and v) a binned scatterplot of the wages variables in the real-world data application.

\printbibliography
%\bibliography{PSEstimator.bib}

\end{document}